# $m^6$A-FORM: An $m^6$A-focused Foundation Model for Decoding $m^6$A Regulatory Function

Ting-He Zhang[1,2], Sumin Jo [2,3], Shou-Jiang Gao[2,4], Yufei Huang [1,2,3,*]

[1]Department of Medicine, University of Pittsburgh School of Medicine, Pittsburgh, PA

[2]Cancer Virology Program, UPMC Hillman Cancer Center, University of Pittsburgh Medical Center, Pittsburgh, PA

[3]Department of Electrical and Computer Engineering, Swanson School of Engineering, University of Pittsburgh, Pittsburgh, PA

[4]Department of Microbiology and Molecular Genetics, University of Pittsburgh School of Medicine, Pittsburgh, PA

*Correspondence: Yufei Huang, Cancer Virology Program, University of Pittsburgh School of Medicine, Pittsburgh, PA 15213, USA. E-mail: YUH119@pitt.edu

## Abstract

N6-methyladenosine ($m^6$A) regulates mRNA fate through site-specific methylation, reader recognition and downstream effects on RNA stability and decay. However, current computational approaches focus mainly on site prediction, leaving unresolved the broader challenge of inferring $m^6$A regulatory context and function from epitranscriptomic profiles. Here we present $m^6$A-FORM, an $m^6$A-focused foundation model for for regulatory discovery. Pretrained on 24.9 million RNA sequence windows from 22.5 million MeRIP-seq regions across 143 human studies, $m^6$A-FORM learns reusable representations of $m^6$A-associated transcript contexts. We adapt this encoder to single-nucleotide $m^6$A discovery, regulator-binding prediction, YTHDF2-associated decay prediction and tissue-scale epitranscriptomic mapping. $m^6$A-FORM predicts binding of 19 $m^6$A readers, writers and erasers and identifies sequence and RBP-context features associated with YTHDF2-mediated RNA degradation. Applied to 67 datasets from 24 human tissues, it identifies tissue-conserved $m^6$A sites linked to stronger methylation, reader binding, RBP occupancy and decay propensity.

# Introduction

N6-methyladenosine ($m^6A$) is the most abundant internal modification in eukaryotic mRNA and a central regulator of post-transcriptional gene expression[1-4]. Its biological effects are not determined by methylation sites alone, but by a regulatory cascade that includes methylation deposition, recognition by $m^6A$-binding proteins, interaction with local RNA-binding protein environments and downstream effects on RNA stability, translation, localization and decay [1, 5-7]. In particular, reader proteins such as YTHDF-family members can confer distinct and sometimes opposing regulatory outcomes, including transcript destabilization or stabilization [8-11]. Dysregulated $m^6A$ programs have been implicated in multiple human diseases, including cancer, where they influence tumor progression, stemness, immune evasion and therapeutic response[12, 13]. Thus, a central goal in $m^6A$ biology is not only to identify where $m^6A$ occurs, but also to determine which regulatory proteins interact with methylated transcripts and which methylation events are likely to alter RNA fate. This requires $m^6A$ profiles that are scalable, site-resolved and functionally informative.

Experimental technologies have mapped the $m^6A$ epitranscriptome, but they capture only parts of the regulatory process. MeRIP-seq, or $m^6A$-seq, remains the most widely used transcriptome-wide profiling method and has enabled large-scale studies across tissues, cell lines and disease contexts[2]. However, it identifies enriched transcript regions shaped by fragmentation, read length and peak calling, rather than the modified adenosine or its functional consequence. Base-resolution technologies, including miCLIP [14] and GLORI [15], provide more precise site information but are more costly, technically demanding or less broadly available at scale. Even when $m^6A$ sites are known, functional interpretation remains difficult. Reader binding depends on local sequence context, site accessibility, neighboring $m^6A$ events and RBP co-occupancy, while $m^6A$-mediated decay depends on whether reader engagement occurs in a regulatory environment that promotes or counteracts transcript destabilization. Therefore, $m^6A$ discovery requires a framework that can connect scalable epitranscriptomic profiling to site-level resolution, regulator binding and functional RNA outcomes.

Computational methods have become indispensable for $m^6A$ annotation, but most existing approaches address only part of this problem. Early $m^6A$ predictors (e.g., SRAMP [16], WHISTLE [17], HS$m^6$AP [18]) used hand-crafted sequence features and conventional machine-learning classifiers, while recent deep-learning models including deepSRAMP [19], MultiRM [20], $m^6A$-TCPred [21], MASS [22], have improved site prediction through representation learning. However, most predictors adopt an adenosine-centered formulation in which each candidate adenosine is represented by a short local window and classified independently. This strategy is computationally inefficient and biologically incomplete. It does not naturally support joint inference among multiple candidate sites within the same transcript region[15], and it does not directly model the broader regulatory context in which $m^6A$ functions. Separate models have also been developed for RNA-protein binding [23, 24] or RNA stability prediction [25], but these are usually not trained in an $m^6A$-aware representation framework and do not explicitly link $m^6A$ sites to $m^6A$ regulator binding and decay-associated regulatory outcomes. As a result, current computational approaches leave a gap between $m^6A$ site prediction and functional discovery.

Recent biological foundation models offer a path toward bridging this gap. Genomic and RNA language models have shown that self-supervised pretraining can learn reusable sequence representations, and recent studies increasingly frame foundation models as systems for biological function prediction rather than single-task classifiers [26-28]. However, general-purpose DNA or RNA models are not necessarily optimal for $m^6A$ biology. Instead, model performance depends critically on whether the pretraining data, input context length, tokenization strategy and fine-tuning formulation are aligned with the biological structure of the downstream problem[29]. This consideration is especially important for $m^6A$, where the relevant structure includes sparse single-base methylation labels, local clustering of candidate sites, context-dependent reader binding, RBP-mediated modulation of RNA fate and tissue-dependent recurrence of regulatory marks. These features create an $m^6A$-specific site-to-function learning problem that differs from generic RNA sequence modeling and from conventional adenosine-centered site prediction.

Here we present $m^6A$-FORM, an $m^6A$-focused Foundation mOdel for decoding $m^6A$ regulatory biology. $m^6A$-FORM is pretrained on MeRIP-seq-derived transcript regions from 143 human studies, using the broad availability of MeRIP-seq data to learn representations from $m^6A$-associated sequence contexts. We then adapt the pretrained encoder to a hierarchy of $m^6A$-focused tasks. For site discovery, $m^6A$-FORM reformulates base-resolution $m^6A$ prediction as regional sequence labeling, enabling joint inference over candidate adenosines within transcript regions. For regulatory discovery, the same representation framework is used to predict binding of $m^6A$ readers, writers and erasers and to identify RBP-context features associated with regulator-specific binding. For functional discovery, $m^6A$-FORM integrates YTHDF2 binding and mRNA half-life information to prioritize $m^6A$ sites associated with RNA degradation. Finally, we apply the model across multi-tissue MeRIP-seq datasets to map tissue-conserved and infrequent $m^6A$ sites and connect recurrent methylation to reader binding, decay propensity and local RBP regulatory environments. Together, these analyses position $m^6A$-FORM as more than a site predictor. It provides a unified representation framework for moving from epitranscriptomic profiling to single-nucleotide discovery, regulator-binding interpretation and functional interpretation of $m^6A$-associated regulatory outcomes By aligning foundation-model pretraining with the biological and experimental structure of $m^6A$ regulation, $m^6A$-FORM enables scalable functional annotation of the human $m^6A$ epitranscriptome.

# Result

## Model overview

To reduce the high false-positive rates that limit existing computational predictors, we developed $m^6A$-FORM, a transformer-based foundation model that learns $m^6A$-enriched sequence representations from large-scale MeRIP-seq data. By aligning model inputs to MeRIP-seq peak regions enriched for methylation, $m^6A$-FORM incorporates an enrichment-defined prior that concentrates candidate sites into biologically relevant sequence space, thereby reducing the search space and improving performance in downstream tasks.

We collected 143 MeRIP-seq studies (Homo sapiens) published through 2024 and processed them using a systematic pipeline with $m^6A$-suite [30]. From the peak calling results, we extracted peak-overlapping transcript sequences to construct the pretraining sequence set (Fig. 1a, Methods). In total, the pretraining dataset comprises 24,909,934 Homo sapiens mRNA sequences derived from 22,548,379 MeRIP-seq peaks, providing broad coverage of methylation-enriched contexts for representation learning. We pretrained m$^6$A-FORM using masked language modeling, allowing the model to learn context-dependent representations of RNA methylation patterns within enriched regions (Fig. 1b).

After pretraining, m$^6$A-FORM serves as an m$^6$A-aware encoder for multiple downstream epitranscriptomic tasks, including single-base m$^6$A site identification, m$^6$A regulator binding site prediction, and m$^6$A sites for YTHDF2-mediated mRNA degradation prediction.

For base-resolution site calling, we fine-tuned a task-specific variant, m$^6$A-FORM-sites, with two innovations that preserve peak-level context while enabling nucleotide-level outputs. First, instead of extracting adenosine-centered windows, we represent the entire sequence within each MeRIP-seq peak as overlapping 3-mers (stride = 1), using the central nucleotide to maintain positional resolution across the peak. Second, we formulate m$^6$A site identification as a sequence-labeling (NER-style) task. Each 3-mer token is labeled 1 if its central adenosine corresponds to an annotated m$^6$A site, 0 if it is unmodified, and -100 (ignored) if the central nucleotide is not an adenosine (Fig. 2b). This formulation enables joint inference over all adenosines within a peak, improving efficiency and allowing the model to capture local dependencies among nearby candidate sites. (Methods)

Beyond base-resolution site calling, m$^6$A-FORM enables downstream tasks such as predicting binding of m$^6$A regulators (writers, erasers, and readers) and identifying m$^6$A sites associated with YTHDF2-mediated mRNA degradation, providing an integrated framework for epitranscriptomic research (Fig. 1c).

## m$^6$A-FORM-sites enables precise single-base m$^6$A prediction

m$^6$A-FORM-sites, a task-specific fine-tuned derivative of m$^6$A-FORM, performs single-base prediction of m$^6$A sites. Because experimentally reported m$^6$A calls can include false positives [31], we constructed a high-confidence, base-resolution set as ground-truth by intersecting sites detected by 13 base-resolution technologies from m$^6$A-Atlas v2.0 and GLORI assay [15, 32]. m$^6$A-Atlas v2.0 aggregates dataset from 12 single-base m$^6$A sequencing methods, whereas GLORI offers antibody-free, base-resolution detection.

We defined high-confidence m$^6$A sites as those sites supported by at least two independent technologies and overlapping with MeRIP-seq peaks (Fig. 1a, Methods); all other m$^6$A sites were excluded. Negative sites were defined as adenosines located within the same MeRIP-seq peaks but not as high-confidence m$^6$A sites. Using these criteria, we built a high-confidence dataset comprising 131,320 single-base m$^6$A sites.

We collected 528,452 MeRIP-seq peaks and partitioned the peaks into training (70%), validation (10%), and test (20%) subsets. Peaks longer than 500 bp were segmented into 500 bp sliding windows with 25-bp overlap to standardize input length. To mitigate class imbalance during optimization, the training and validation sets were built using a 1:10 positive-to-negative ratio achieved by random sampling of non-m$^6$A adenosines, whereas the test set preserved the natural class distribution to reflect realistic conditions (Methods). To ensure robustness to data splitting and sampling variability, we repeated training and evaluation five times using different random seeds to partition peaks and report combined performance.

We benchmarked m$^6$A-FORM-sites against state-of-the-art m$^6$A predictors. We also included an architectural control, m$^6$A-FORM–baseline, which uses the same model and fine-tuning structure but is trained from scratch without m$^6$A-FORM pretraining weights. m$^6$A-FORM-sites achieved state-of-the-art performance for $m^6A$ site prediction, with a mean precision–recall AUC (PR-AUC) of 0.635 and ROC-AUC of 0.988 (Figure 2c,e). This yields at least a 0.14 PR-AUC improvement over the best prior model (DeepSRAMP), underscoring more precision in identifying true $m^6A$ sites under realistic class imbalance. Notably, while $m^6A$-FORM-sites and the baseline attain similar ROC–AUC, the pretraining step yields a substantial improvement in PR–AUC. This suggests that the model achieves improved precision under strong class imbalance and more reliably prioritizes true $m^6A$ sites in peak contexts.

To further dissect the contribution of model pretraining and the NER-style labeling strategy, we performed an ablation analysis by comparing $m^6A$-FORM-sites with variants using A-centered labeling or without pretraining (Methods). The full $m^6A$-FORM-sites model achieved the best overall performance, with both high PR-AUC and ROC-AUC, whereas removing pretraining or replacing the NER-style labeling strategy led to reduced performance (Fig. 2d). These results indicate that large-scale pretraining enables the model to capture informative sequence representations associated with $m^6A$, while the NER-style labeling strategy further improves site-level prediction by allowing simultaneous evaluation of all candidate adenosines within an input sequence. This formulation also substantially improved inference efficiency. By predicting all candidate adenosines in a sequence simultaneously, $m^6A$-FORM-sites avoided repeated A-centered window inference and achieved an approximately 10.7-fold faster GPU runtime than the A-centered labeling method (Fig. 2f).

Finally, we examined how label coverage influenced model performance by progressively relaxing the ground-truth inclusion criteria. We considered three criteria: (i) sites supported by $\geqslant 2$ technologies within $m^6A$-Atlas v2.0; (ii) sites supported by $\geqslant 2$ technologies when considering $m^6A$-Atlas v2.0 and GLORI jointly (current setting), and (iii) all sites reported by $m^6A$-Atlas v2.0 and GLORI. Relaxing these criteria increased the number of labeled positives and yielded monotonic gains in PR-AUC (Fig. 2g), indicating that $m^6A$-FORM-sites could benefit from larger training sets while remaining sensitive to the precision–coverage trade-off imposed by label confidence.

## $m^6A$-FORM-RWEBind improves prediction of $m^6A$ regulator binding sites

$m^6A$ regulators are RNA-binding proteins (RBPs) that install (“writers”), remove (“erasers”), or recognize (“readers”) $m^6A$ modification, thereby influencing RNA stability, splicing, export, and translation[33]. Accurately predicting $m^6A$–regulator interactions is important for understanding post-transcriptional regulation and their impact on gene expression.

We collected publicly available CLIP–seq datasets for known $m^6A$ regulators (Readers, Writers, Erasers; RWEs) from POSTAR3, which provides high-confidence RBP binding peaks [34]. This yielded a comprehensive RWE binding dataset comprising 104,244 sequences across 19 RWEs (13 readers, 5 writers, and 1 eraser). These CLIP peaks were intersected with our high-confidence single-base $m^6A$ sites. For each RWE, an $m^6A$ site overlapping its CLIP peak was labeled positive (label = 1), whereas sites without overlap were labeled negative (label = 0). Training and validation used a 1:10 positive-to-negative ratio via random down-sampling of negatives, while the test set preserved the natural class distribution.

Because regulator-binding labels are sparse and site-specific[14], NER–style sequence-labeling formulations can be sensitive to label sparsity and limited positive supervision [35]. We therefore formulated RWE binding prediction as a fixed, $m^6A$-centered sequence classification task. For each candidate $m^6A$ site, we extracted a ±50-bp window and fine-tuned $m^6A$-FORM to predict regulator binding using the pretrained methylation-aware encoder (Methods). This formulation accommodates sparse binding labels while retaining the $m^6A$-enriched representations learned from MeRIP-seq pretraining.

We benchmarked $m^6A$-FORM-RWEBind against widely used RBP-binding predictors, including iDeepS [23] and RNAProt [24]. Across most RWEs, $m^6A$-FORM-RWEBind consistently outperformed these benchmark models, with especially strong gains for regulators with larger training sets. This is consistent with the model's ability to improve utilization of richer training signal as data availability increases (Supplementary). Overall, $m^6A$-FORM-RWEBind achieved a median PR–AUC improvement of 0.09, with significant gains for 15 of 19 RWEs. These results suggest that $m^6A$-aware representations learned during pretraining improve regulator binding prediction relative to models that do not explicitly encode epitranscriptomic context.

We next examined whether regulator-specific prediction performance was influenced by the size of the available training data. Across the 19 $m^6A$ regulators, PR-AUC showed a positive correlation with training-set size ($R = 0.75$; Fig. 3e), indicating that regulators with more CLIP-supported training examples generally achieved better predictive performance.

To interpret sequence and contextual features driving predictions, we computed Integrated Gradients attribution scores for representative regulators (YTHDF2, YTHDC1, IGF2BP1, and RBM15) and identified high-attribution subregions within input sequences (Fig. 3b, Methods). We then tested whether CLIP peaks of other RBPs were preferentially enriched within these high-attribution subregions between the positive and negative test set using Fisher's exact test (Methods; Supplementary).

This analysis revealed regulator-specific RBP neighborhoods with strong enrichment signals across multiple functionally related factors (Fig. 3f). RBM15-associated regions were enriched for RBM15B and additional RNA-processing factors, reflecting the close functional relationship between RBM15-family proteins and the $m^6A$ methyltransferase-associated machinery [36]. YTHDC1-associated regions showed enrichment of multiple nuclear RNA-processing and splicing-related RBPs, including SRSF-family proteins and hnRNP factors, consistent with the established role of YTHDC1 in nuclear mRNA processing [37]. Additionally, YTHDF2-associated regions were enriched for YTHDF-family proteins, including YTHDF1 and YTHDF3, together with decay- and surveillance-associated factors such as UPF1. The co-enrichment of all three YTHDF proteins is consistent with the proposed unified model in which YTHDF readers can act cooperatively or redundantly on $m^6A$-modified transcripts[11], whereas UPF1 enrichment is consistent with reported links between UPF1 and YTHDF2-associated decay of $m^6A$-containing RNAs [38]. IGF2BP1-associated regions were strongly enriched for IGF2BP-family members, consistent with shared stabilizing programs among IGF2BP paralogs [39]. Co-enrichment of IGF2BP1 and YTHDF2 suggests that these transcripts may be subject to opposing $m^6A$-linked regulations outcomes (decay versus stabilization) [40, 41], motivating further analysis of context-dependent YTHDF2-associated decay. These regulator-specific RBP neighborhoods are consistent with known functional distinctions among $m^6A$ regulators and suggest that $m^6A$ regulatory outcomes are shaped not only by individual $m^6A$ readers or writers, but also by the surrounding RBP context. The recovery of known regulator-specific co-binding patterns supports the biological relevance of the inferred RBP neighborhoods and provides a framework for interpreting context-dependent $m^6A$ regulation.

## $m^6A$-FORM-decay can identify YTHDF2-$m^6A$ associated mRNA degradation

YTHDF2 plays a crucial role in regulating mRNA stability. Accurately predicting and interpreting YTHDF2-mediated decay can provide valuable insights into post-transcriptional regulation. We therefore developed $m^6A$-FORM-decay, a fine-tuned model derived from $m^6A$-FORM to predict $m^6A$ sites associated with YTHDF2-mediated decay and to provide sequence-level interpretability.

To obtain training data, we intersected 131,320 high-confidence $m^6A$ sites with 52,792 YTHDF2 CLIP–seq peaks in HeLa cells and further required overlap with corresponding MeRIP-seq peaks, yielding 207 high-confidence YTHDF2-bound MeRIP-seq peak regions. These peaks were split into training (70%), validation (10%), and test (20%) sets by peak index. We then integrated mRNA half-life measurements[1] to assign decay labels. $m^6A$ sites within YTHDF2 Par-CLIP peaks and located in genes exhibiting increased half-life under YTHDF2 knockdown were labeled as positive and $m^6A$ sites within YTHDF2 Par-CLIP peaks but without increased half-life were labeled as negative (Fig. 4a, Methods). For each $m^6A$ site within these peaks, we extracted A-centered windows (±250 bp), yielding 582 positive and 6,700 negative sequences. To mitigate class imbalance, the training and validation sets were randomly downsampled to a 1:1 positive-to-negative ratio, while the test set retained the natural distribution. We repeated peak partitioning and model training and evaluation five times using different random seeds (Methods).

We compared $m^6A$-FORM-decay with $m^6A$BERT-decay (Fig. 4b). Although $m^6A$BERT-decay achieved a slightly higher ROC-AUC (0.730 vs 0.703), $m^6A$-FORM-decay achieved a higher PR-AUC (0.167 vs 0.150). Given the pronounced class imbalance, the PR-AUC improvement indicates that $m^6A$-FORM-decay provides a better precision–recall trade-off and more effectively prioritizes $m^6A$ sites associated with YTHDF2-mediated mRNA decay.

To elucidate the mechanistic basis of mRNA decay, we applied Integrated Gradients to compute nucleotide-level attribution scores (Methods). These scores quantify the contribution of each input nucleotide to the model prediction, with higher scores indicating stronger contribution. Visualization of attribution profiles revealed distinct patterns between decay-associated and non-decay sites (Fig. 4c,d). Positive sequences showed relatively diffuse and low attribution across the A-centered window, whereas negative sequences exhibited more structured attribution signals. Hierarchical clustering of the negative set identified two groups: negative G1, comprising 1,616 sequences with a prominent attribution peak approximately 70 bp upstream of the $m^6A$ site, and negative G2, comprising 3,755 sequences with more distributed attribution peaks, including signals near approximately 70 bp upstream and 20 bp downstream of the $m^6A$ site.

We next identified contiguous high-attribution subregions (Methods) and mapped known HeLa RBP binding sites from POSTAR3 onto these regions. For each RBP, we quantified the fraction of high-attribution regions overlapping its binding sites and compared enrichment between positive and negative groups using two-sided Fisher's exact tests followed by Benjamini–Hochberg correction. To focus on robust and interpretable effects, we further required enriched RBPs to pass both statistical and effect-size filters: $FDR<0.05$, $|\log_2(\text{odds ratio})|>\log_2(1.5)$ and $|\text{odds ratio difference}|>0.01$ (Methods).

Using these criteria, we identified eight RBPs showing significant differential enrichment between positive and negative high-attribution regions (Fig. 4e), including HNRNPC, CSTF2, TIAL1, ACI, U2AF65, TIA1, PTBP1, and ALYREF. All eight RBPs were preferentially enriched in negative high-attribution regions. Many of these factors have been associated with transcript stabilization, whereas others may represent newly identified stabilizing factors[42-46]. This suggests that non-decay YTHDF2-bound m$^6$A sites may be associated with stabilizing or protective RBP contexts rather than YTHDF2 binding alone.

We further performed cluster-stratified enrichment analyses to examine whether the two negative clusters represented distinct non-decay regulatory contexts. When positive high-attribution regions were compared with each negative cluster separately, several RBPs were recurrently identified across comparisons, including HNRNPC, CSTF2, TIAL1, ACI, TIA1, PTBP1, and ALYREF, suggesting that these RBP-context features distinguish decay-associated sites from multiple non-decay backgrounds. Direct comparison between negative G1 and negative G2 further identified HNRNPC, CSTF2, and TIA1 as differentially enriched between the two non-decay clusters, indicating intrinsic heterogeneity among non-decay YTHDF2-bound m$^6$A sites. Together, these results suggest that the regulatory outcome of YTHDF2-bound m$^6$A sites is shaped by local RBP context. Non-decay sites were preferentially associated with putative stabilizing or protective RBP environments, suggesting that local RBP environments may modulate whether YTHDF2-bound transcripts undergo degradation.

## m$^6$A-FORM maps a multi-tissue human m$^6$A landscape and identifies tissue-conserved sites linked to expression regulation

To build a robust multi-tissue landscape of human m$^6$A, we processed 67 MeRIP-seq datasets from 24 tissues using the m$^6$A-suite[30] pipeline. These datasets comprised 22 samples from five male donors in the Chinese Brain Bank Center (CBBC) [47] and 45 samples from seven donors from the National Genomics Data Center (NGDC) dataset [48]. Using filtered peak sets from each sample (FDR $< 0.05$; see Methods), m$^6$A-FORM identified 157,112 predicted m$^6$A sites in CBBC and 320,052 in NGDC (Fig. 5a).

Although peak counts varied substantially across datasets, likely reflecting differences in sequencing depth and experimental protocols, the average number of predicted sites per peak was highly consistent between CBBC and NGDC (Fig. 5c). This consistency suggests that m$^6$A-FORM-sites produced stable site-level predictions despite dataset-level differences in peak abundance.

After combining all predictions across samples and collapsing redundant calls (Methods), we identified 317,986 non-redundant m$^6$A sites. These sites were primarily distributed across CDS (45.8%) and 3′UTRs (38.67%), with only limited localization to 5′UTRs (7.33%), broadly consistent with reported m$^6$A metagene distributions (Fig. 5b). Consistently, metagene analysis across 24 human tissues revealed a highly conserved positional distribution of predicted m$^6$A sites along transcripts (Fig. 5e).

Although m$^6$A-FORM identified a large repertoire of predicted m$^6$A sites across individual samples, these sites varied substantially in their recurrence across tissues and cohorts: some sites were repeatedly detected across multiple biological contexts, whereas many others appeared only sporadically. We therefore asked whether a statistically supported core set of recurrent m$^6$A sites could be distinguished from infrequent or tissue-restricted events. To address this, we quantified sample recurrence for each predicted site and compared it with permutation-based null distributions separately in the CBBC and NGDC datasets (Fig. 5d). In both datasets, observed $m^6A$ sites recurred across more samples than expected by chance, indicating non-random recurrence across tissues and individuals. Sites showing significant recurrence in both datasets after FDR correction were defined as tissue-conserved (TC) $m^6A$ sites (FDR < 0.05). This analysis identified 19,631 TC sites, representing a compact and reproducible set of recurrent $m^6A$ sites across diverse biological contexts. In contrast, 241,000 sites were classified as infrequent $m^6A$ sites because they were detected in no more than three tissues. Sites with intermediate recurrence were not included in the subsequent TC-versus-infrequent comparison. These definitions provided the basis for downstream characterization of TC and infrequent $m^6A$ regulatory features (Fig. 6a).

We next asked whether TC sites exhibit distinct signatures compared with infrequent sites. Attribution analysis in local sequence windows (±5 bp) revealed substantially higher attribution scores around TC sites than around infrequent sites across 24 tissues (Fig. 6e), suggesting that TC sites are supported by more consistent sequence-context features across tissues.

Recent GLORI-based measurements, which chemically distinguish methylated from unmethylated adenosines, revealed that $m^6A$ sites frequently occur in clusters[15]. Consistent with this, TC sites were significantly enriched in GLORI-defined clustered regions compared to infrequent sites. Among 19,631 TC $m^6A$ sites and 241,000 infrequent $m^6A$ sites, overlaps were more common in clustered than non-clustered regions (10,094 vs. 5,369), whereas infrequent sites showed the opposite pattern (16,048 vs 30,925) among sites mappable to GLORI-defined regions (two-sided Fisher's exact test; odds ratio = 3.62, $P < 2.2 \times 10^{-16}$).

Compared with infrequent sites, TC sites showed significantly stronger enrichment near stop codons, a hallmark of canonical $m^6A$ topology (Fig. 6b). While $m^6A$ sites generally localize to 3′ UTRs, TC sites showed a sharper density peak, indicating preferential localization to this regulatory region. We then compared methylation levels across site groups and found that TC sites exhibited significantly higher methylation than infrequent sites (Wilcoxon test, $P < 2.2 \text{x} 10^{-16}$; Fig. 6g), a pattern consistently observed across samples (Supplementary Fig. SX). Because methylation level is expected to reflect the degree of regulatory engagement at a site, we next examined whether methylation levels at TC sites were associated with gene expression. Integrating matched expression profiles, we observed a negative correlation between TC-site methylation levels and gene expression (Spearman correlation, -0.45, $P < 2.2 \times 10^{-16}$). Together, these results suggest that TC methylation is linked to reduced transcript abundance, consistent with $m^6A$'s known role in post-transcriptional regulation, and motivate mechanistic analyses focused on decay-associated reader activity.

Given the central role of YTHDF2 in $m^6A$-related mRNA decay, we tested whether TC sites preferentially exhibit YTHDF2-preferred binding and decay-associated signatures. Using $m^6A$-FORM-RWEBind, we predicted YTHDF2 binding preference at $m^6A$ sites. Among 67,309 sites predicted to favor YTHDF2 binding, 6,128 of 19,631 TC sites were predicted as YTHDF2-preferred, compared with 40,603 of 241,000 infrequent sites. TC sites were significantly enriched for predicted YTHDF2 binding (Fisher's exact test; $P < 2.2 \times 10^{-16}$, odds ratio = 2.34). To further connect binding potential to functional outcome, we applied $m^6A$-FORM-decay to predicted YTHDF2-associated sites to infer decay propensity. We found 5,546 of 6,128 TC sites associated with decay signatures, compared to 35,823 among 40,603 infrequent sites (two-sided Fisher's exact test; $P < 1.07 \times 10^{-7}$, odds ratio = 1.27). These results suggest that TC $m^6A$ sites are preferentially associated with YTHDF2 binding and predicted decay propensity, supporting a potential conserved role in post-transcriptional destabilization.

We next examined whether TC sites are embedded in distinct local RBP regulatory environments. Compared with infrequent sites, TC sites overlapped a significantly larger number of unique RBP binding events per site among both YTHDF2-associated sites and YTHDF2-decay-associated sites (Fig. 6h,i). This suggests that TC $m^6A$ sites are located in RBP-rich regulatory neighborhoods, consistent with the idea that conserved methylation marks are embedded within broader post-transcriptional regulatory modules.

To characterize the functional relevance of these RBP contexts, we performed enrichment analysis on RBPs preferentially associated with TC sites predicted to bind YTHDF2. Gene Ontology Biological Process analysis revealed enrichment of RNA metabolic processes, RNA stability regulation, and RNA catabolic processes (Fig. 6j), supporting a link between TC-associated RBP environments and post-transcriptional regulation. We further focused on YTHDF2-decay-associated sites and compared RBP enrichment between TC and infrequent sites. This analysis identified several RBPs enriched near TC decay-associated sites, including HNRNPC, UPF1, YTHDF1, YTHDF3, CPSF6, CPSF7, and ELAVL1 (Fig. 6k). Many of these RBPs have been implicated in RNA stability and processing. These results indicate that TC $m^6A$ sites are coupled to local RBP environments linked to RNA processing and stability, providing a potential context through which conserved $m^6A$ marks may contribute to transcript destabilization.

# Discussion

In this study, we present $m^6A$-FORM, an $m^6A$-focused foundation model that addresses the gap between scalable MeRIP-seq profiling and functional interpretation of $m^6A$ regulation by connecting peak-derived pretraining, single-base site discovery, regulator binding, YTHDF2-associated decay prediction and tissue recurrence analysis. By incorporating MeRIP-seq-derived methylation-enriched regions as an $m^6A$-specific biological prior, $m^6A$-FORM learns representations from approximately 24.9 million RNA sequence windows derived from 22.5 million MeRIP-seq peaks across 143 human studies. This design links large-scale self-supervised pretraining with the peak-to-site structure of $m^6A$ discovery, distinguishing $m^6A$-FORM from both adenosine-centered predictors and general-purpose RNA foundation models trained on generic sequence corpora.

More broadly, $m^6A$-FORM supports the emerging view that effective biological foundation models should be matched to the structure of the target problem rather than defined only by model scale or generic sequence pretraining. Recent benchmarking studies have shown that RNA and nucleotide language models are not universally optimal across all RNA-related tasks, and that performance depends on the alignment between pretraining data, input context, tokenization strategy and fine-tuning formulation [29]. For $m^6A$ biology, the relevant structure is defined by peak-level MeRIP-seq enrichment, sparse single-base methylation labels, local clustering of candidate sites, regulator and RBP context, and tissue-dependent recurrence. By matching the pretraining corpus, input representation and fine-tuning formulations to these $m^6A$-specific properties, $m^6A$-FORM demonstrates the value of specialized foundation models for epitranscriptomic regulation.

Within this framework, base-resolution site discovery represents the first step toward functional interpretation, where the central challenge is to convert scalable but region-level MeRIP-seq signals into precise single-nucleotide annotations. The performance of $m^6A$-FORM-sites highlights the importance of modeling $m^6A$ sites within their peak-level regulatory context. Existing computational predictors commonly evaluate short fixed-length windows around individual adenosines independently, even though true $m^6A$ sites represent only a small fraction of transcriptomic adenosines and often occur within methylation-enriched regions. This adenosine-centered formulation can lead to redundant inference and an increased false-positive burden, even when motif constraints are used to restrict candidate sites [17]. In contrast, $m^6A$-FORM-sites reformulates base-resolution $m^6A$ discovery as a sequence-labeling task within MeRIP-seq peak-derived regions, enabling joint prediction over all candidate adenosines in a local transcript context. The improvement over adenosine-centered and non-pretrained baselines indicates that both methylation-enriched pretraining and peak-level sequence labeling contribute to more accurate, efficient and context-aware $m^6A$ site prediction. This formulation also improves computational scalability by avoiding repeated inference over overlapping adenosine-centered windows, making it more suitable for large-scale epitranscriptomic mapping.

Beyond site calling, $m^6A$-FORM extends the analysis of $m^6A$ from where methylation occurs to which regulatory proteins interact with methylated transcripts and what RNA-fate outcomes these interactions may produce. As a shared $m^6A$-aware representation framework, $m^6A$-FORM supports task-specific adaptation for predicting binding sites of 19 $m^6A$-associated regulators, indicating that methylation-enriched pretraining captures sequence features relevant to $m^6A$-associated protein recognition. Attribution-guided analysis further identified regulator-specific RBP neighborhoods, including co-enrichment patterns consistent with known relationships among RBM15-family proteins, YTHDC1-associated nuclear RNA-processing factors, YTHDF-family readers and IGF2BP-family stabilizing factors. These findings suggest that $m^6A$ regulator binding is shaped not only by the methylated site itself, but also by local RBP environments that may modulate downstream regulatory outcomes.

Among these regulatory outcomes, YTHDF2-associated decay provides a direct test case for linking $m^6A$ reader engagement to RNA fate. By integrating YTHDF2 binding information with mRNA half-life changes, $m^6A$-FORM-decay prioritized $m^6A$ sites associated with YTHDF2-mediated degradation and revealed distinct attribution patterns between decay-associated and non-decay sites. The interpretation of non-decay sites suggests that stabilizing or protective RBP contexts may counteract or modify YTHDF2-associated decay. These results support a context-dependent model of $m^6A$ function, in which the regulatory consequence of a methylated transcript depends on the local balance among decay-promoting readers, stabilizing factors and other RNA-binding proteins.

The multi-tissue application of $m^6A$-FORM demonstrates how scalable site-level inference can be used to investigate recurrent regulatory methylation across human tissues. By applying $m^6A$-FORM-sites to 67 MeRIP-seq datasets from 24 tissues, we generated a single-base $m^6A$ landscape and identified a compact set of tissue-conserved $m^6A$ sites supported by recurrence analysis in two independent multi-tissue cohorts. Compared with infrequent sites, tissue-conserved sites showed stronger canonical $m^6A$ topology, higher methylation levels, stronger model attribution, enrichment in GLORI-defined clustered regions, increased predicted YTHDF2 binding and decay propensity, and greater local RBP occupancy. These features suggest that tissue-conserved sites represent a recurrent regulatory layer of the human $m^6A$ epitranscriptome, whereas infrequent sites may reflect more context-dependent, condition-specific or lower-confidence methylation events.

These observations connect recurrence of methylation across tissues with potential post-transcriptional regulatory function. The enrichment of RNA metabolism, RNA stability and RNA catabolic processes among RBPs associated with tissue-conserved YTHDF2-related sites further supports the idea that recurrent m$^6$A marks are embedded within broader RNA regulatory modules. Thus, the tissue-scale analysis provides an example of how m$^6$A-FORM can move beyond cataloging predicted sites to prioritizing methylation events with potential regulatory consequences.

Several limitations define directions for future work. First, m$^6$A-FORM relies on MeRIP-seq-derived peak regions for pretraining and inference. This strategy provides a useful methylation-enriched biological prior and enables large-scale learning from widely available datasets, but it may also inherit biases from antibody-based enrichment, peak-calling pipelines, sequencing depth, tissue coverage and experimental conditions. Future work incorporating additional antibody-free and base-resolution datasets may further improve the robustness and generalizability of the learned representations. Second, the high-confidence training labels were constructed from sites supported by m$^6$A-Atlas v2.0 and GLORI to prioritize label precision. This strategy reduces noise in positive labels but excludes sites supported by only one technology, potentially introducing mislabeled positives into the negative set and limiting sensitivity to low-signal or condition-specific methylation sites. Future versions could incorporate soft labels, uncertainty-aware learning or technology-specific confidence scores. Third, m$^6$A-FORM currently relies primarily on sequence information and MeRIP-seq-informed pretraining. Incorporating additional modalities, such as RNA structure, accessibility, expression level, cell type, subcellular localization and perturbation context, may further improve functional interpretation of m$^6$A sites. Finally, broader comparisons with existing RNA and nucleotide foundation models under matched peak-to-site and site-to-function settings will be valuable for further defining when m$^6$A-specific pretraining provides the greatest advantage.

In summary, m$^6$A-FORM provides a unified m$^6$A-aware foundation-model framework for moving from MeRIP-seq peaks to single-base sites, regulator-binding prediction, RBP-context interpretation, YTHDF2-associated functional annotation and tissue-scale regulatory discovery. By aligning methylation-enriched pretraining and task-specific adaptation with the biological and experimental structure of m$^6$A regulation, m$^6$A-FORM moves beyond conventional site prediction and offers a practical resource for decoding the regulatory context and functional consequences of the human m$^6$A epitranscriptome.

# Methods

## Pretraining data preparation

We collected 143 Homo sapiens MeRIP-seq samples from GEO, with data available through March 2024. Raw FASTQ files were downloaded using SRA Toolkit and processed using the m$^6$A-suite pipeline [30]. Briefly, read quality control was performed using fastp [49], and adapter sequences were trimmed using cutadapt [50] v2.10 with the parameters --minimum-length 25 --pair-filter=any. Trimmed reads were aligned to the GRCh38 genome assembly using HISAT2 v2.2.1 [51] with the parameters --no-mixed --no-discordant. For each sample, aligned IP and input BAM files were used for m$^6$A peak calling with exomePeak2 [52] in transcriptome mode. High-quality peaks were selected based on the following criteria: (i) FDR < 0.05; (ii) assignment to a single gene with detectable expression; (iii) non-zero peak read counts in both IP and input samples; and (iv) peak length < 1,000 bp. This procedure yielded 22,548,379 peak regions with a mean length of 219 bp.

For each peak, we extracted the corresponding RNA sequence. Because our model input length is limited up to 512 nt, peak sequences were segmented into 500-nt windows using a sliding-window strategy with 25-nt overlap to preserve context information and avoid truncating potential $m^6A$ sites near window boundaries. This procedure produced 24,909,934 mRNA sequence windows, forming a large-scale pretraining corpus for learning $m^6A$-associated sequence patterns.

## Ground truth construction for $m^6A$

To obtain a base-resolution ground truth for $m^6A$ site identification, we integrated single-nucleotide $m^6A$ sites from $m^6A$-Atlas v2.0 and GLORI[15, 32]. $m^6A$-Atlas v2.0 aggregates datasets from 12 single-based $m^6A$ profiling technologies, whereas GLORI provides antibody-free, base-resolution detection. We harmonized all sites to a common reference coordinate system and restricted candidate positions to adenosines located within MeRIP-seq peak regions.

High-confidence ground-truth $m^6A$ sites were defined as sites that overlapped MeRIP-seq peaks and were supported by at least two independent base-resolution technologies. Adenosines within the same MeRIP-seq peaks that were not identified by any base-resolution technology and were not included in the high-confidence positive set were labeled as negatives, while sites supported by only one technology were excluded to reduce label noise. Because miCLIP [14] and miCLIP2 [31] rely on highly similar experimental principles, we treated them as a single technology when counting supporting evidence. Using these criteria, we constructed a high-confidence dataset containing 131,320 base-resolution $m^6A$ sites.

## Dataset preparation for $m^6A$ sites identification

We collected 528,452 MeRIP-seq peaks from five human cell lines with the largest number of available experiments (HeLa, A549, HEK293T, HepG2, and MOLM13). Only untreated/control samples were included. Peaks shorter than 25 bp were excluded to ensure sufficient sequence context. The remaining peaks were randomly split into training (70%), validation (10%), and test (20%) subsets. After partitioning, peaks longer than 500-nt were segmented into 500-nt windows using a sliding-window scheme with 25-nt overlap.

## $m^6A$-FORM architecture and masked language model pretraining

$m^6A$-FORM was implemented as a BERT-like transformer encoder for RNA sequence representation learning. Following the DNABERT [53, 54] framework, each RNA sequence was tokenized into overlapping 3-mers with a stride of one nucleotide. This k-mer tokenization allows the model to capture local RNA sequence patterns while preserving positional information across methylation-enriched transcript regions. Tokenized peak-derived sequences were processed in a line-by-line manner, with each sequence window treated as an independent training instance. The maximum input length was set to 512 tokens, including special tokens.

The model was initialized from a BERT configuration adapted for 3-mer RNA/DNA tokenization and trained using a masked language modeling objective. The encoder consisted of token embeddings, positional embeddings, segment embeddings, stacked multi-head self-attention layers, and feed-forward transformer layers (Fig. 1b). A masked language modeling head was added on top of the final hidden states to predict the identities of masked 3-mer tokens.

To reduce information leakage caused by the strong overlap between adjacent k-mers, we used a span-aware k-mer masking strategy rather than masking isolated tokens only. For each selected 3-mer token, the immediately upstream and downstream neighboring tokens were also masked when they fell within the valid sequence range. This strategy prevents the model from trivially reconstructing a masked k-mer from nearly identical adjacent k-mers and encourages learning broader contextual dependencies within MeRIP-seq peak-derived sequences.

For masked positions, 80% of tokens were replaced with the special `[MASK]` token, 10% were replaced with randomly sampled vocabulary tokens, and the remaining 10% were kept unchanged. Cross-entropy loss was computed only for masked tokens, while unmasked tokens and padding tokens were ignored. No next-sentence prediction objective was used.

Pretraining was performed using AdamW optimization with a learning rate of $1\times10^{-4}$, weight decay of 0.01, Adam epsilon of $1\times 10^{-6}$, $\beta_1$= 0.9, and $\beta_2$ = 0.98. Training used a per-GPU batch size of 50 with gradient accumulation over 25 steps. The model was trained for a maximum of 120,000 optimization steps with 2,000 warmup steps and a linear learning-rate schedule. During pretraining, candidate tokens were randomly selected with a masking probability of 0.15 for the first 100,000 steps, which was increased to 0.25 for the following 20,000 steps. Model checkpoints were saved every 200 optimization steps, and validation perplexity was evaluated during training. The resulting pretrained encoder was used to initialize downstream task-specific models for base-resolution $m^6A$ site identification, $m^6A$ regulator binding prediction, and YTHDF2-mediated mRNA decay prediction. All model training and finetuning was performed on NVIDIA L40S and A100 40GB GPUs.

## Fine-tuning and evaluation of $m^6A$-FORM-sites

To fine-tune $m^6A$-FORM for base-resolution $m^6A$ site identification, we formulated the task as a token-level sequence-labeling (NER-style) problem within MeRIP-seq peak-derived transcript regions. Similar to the pretraining stage, each MeRIP-seq peak-derived sequence was tokenized into overlapping 3-mer tokens with a stride of one nucleotide and used as model input.

For a 3-mer token spanning positions (i, i+1, i+2), the token-level label was assigned according to the central nucleotide at position (i+1). Because labels were assigned to the central nucleotide of each 3-mer, only positions that could serve as the center of a 3-mer were considered for token-level supervision; terminal nucleotides at the beginning and end of each sequence that could not be represented as central positions were ignored. If the central nucleotide was an adenosine corresponding to a high-confidence $m^6A$ site, the token was labeled as positive (1). If the central nucleotide was an adenosine located within a MeRIP-seq peak but not annotated as a high-confidence $m^6A$ site, the token was labeled as negative (0). Tokens whose central nucleotide was not an adenosine were assigned an ignore label (-100) and were excluded from loss calculation. Special tokens introduced by the tokenizer, including classification ([CLS]), separator ([SEP]), and padding ([PAD]) tokens, were also assigned -100.

To reduce class imbalance during model optimization, negative adenosines in the training and validation sets were randomly downsampled to obtain an approximately 1:10 positive-to-negative ratio. Adenosines that were not selected during negative downsampling were assigned -100 and were ignored during loss calculation. In contrast, the test set retained the natural positive-to-negative distribution of adenosines within MeRIP-seq peaks to reflect realistic inference conditions.

Data splitting was performed at the peak level. MeRIP-seq peaks were split into training, validation, and test sets at a ratio of 70:10:20, ensuring that windows derived from the same peak were assigned to the same subset and preventing peak-level data leakage. Peaks longer than 500 bp were divided into 500-bp sliding windows with a 25-bp overlap. This windowing strategy allowed long peak regions to be processed by the transformer while retaining local context around candidate sites.

The pretrained m$^6$A-FORM encoder was fine-tuned with a token-classification head that predicts a binary label for each valid adenosine-centered token. The model was optimized using cross-entropy loss computed only over valid labeled adenosine tokens. Positions labeled as -100 were ignored by the loss function. After training, best model selection was performed using validation-set performance, with the best checkpoint selected according to validation PR-AUC.

For inference, each MeRIP-seq peak-derived sequence or sliding window was processed by the fine-tuned model, and predicted probabilities were extracted only for tokens whose central nucleotide was an adenosine. These token-level probabilities were mapped back to their corresponding genomic coordinates. When the same adenosine was covered by multiple overlapping windows, its final prediction score was obtained by averaging the predicted probabilities across all windows containing that site. A site was called as predicted high confidence m$^6$A if its probability exceeded the selected decision threshold. For benchmark evaluation, threshold-dependent metrics were computed using a decision threshold selected on the validation set. For downstream high-confidence site calling, a stringent probability cutoff of 0.78 was used in this study.

All evaluation metrics were computed at the site level rather than the token-window level. After predictions from overlapping windows were merged, each unique adenosine coordinate in the test set was represented by a single prediction score and a single ground-truth label, ensuring that sites covered by multiple windows were evaluated only once. PR-AUC and ROC-AUC were computed using continuous prediction scores. To assess robustness to data partitioning and negative sampling, the entire fine-tuning and evaluation procedure was repeated using five random seeds. Performance was summarized as the mean and standard deviation across the five independent runs.

## Benchmark models and ablation analyses for m$^6$A identification

To evaluate the performance of m$^6$A-FORM-sites, we compared it with representative state-of-the-art computational predictors for m$^6$A site identification, including DeepSRAMP [55], MultiRM [20], m$^6$A-TCPred [21], and MASS [22]. For each benchmark model, input sequences were prepared according to the input format required by the corresponding method. Because most existing predictors use an adenosine-centered formulation, fixed-length windows centered on candidate adenosines were extracted from the same MeRIP-seq peak-derived regions used for m$^6$A-FORM-sites. Positive and negative labels were defined using the same high-confidence m$^6$A annotations and candidate negative adenosines described above. To ensure a fair comparison, benchmark models were evaluated on the same peak-level test sets as m$^6$A-FORM-sites, and the test set retained the natural positive-to-negative distribution of candidate adenosines.

For benchmark evaluation, prediction scores from all benchmark models were evaluated at the site level. Because the test set retained a strong natural positive-to-negative imbalance, benchmark comparison focused on threshold-independent metrics, including PR-AUC and ROC-AUC. Both PR-AUC and ROC-AUC were computed using continuous prediction scores. The entire training and evaluation procedure was repeated using five random seeds, and performance was summarized as the mean and standard deviation across the five independent runs.

To assess the contributions of pretraining and the NER-style sequence-labeling strategy, we performed a 2 × 2 ablation analysis using four model settings: m$^6$A-FORM-sites, m$^6$A-FORM-sites w/o pretraining, A-centered m$^6$A-FORM-sites, and A-centered m$^6$A-FORM-sites w/o pretraining (Fig. 2d). m$^6$A-FORM-sites represents the full model, using pretrained m$^6$A-FORM weights and the NER-style token-level sequence-labeling formulation. m$^6$A-FORM-sites w/o pretraining used the same transformer architecture, 3-mer tokenization, NER-style token-label construction, peak-level data split, loss function, class-balancing strategy, model-selection criterion, and site-level evaluation protocol as m$^6$A-FORM-sites, but initialized the encoder from scratch rather than from the pretrained m$^6$A-FORM weights. This comparison was used to quantify the contribution of large-scale self-supervised pretraining on MeRIP-seq peak-derived sequences. A-centered m$^6$A-FORM-sites used the pretrained m$^6$A-FORM encoder but replaced the NER-style formulation with an adenosine-centered formulation, in which each candidate adenosine was represented by a fixed-length sequence window centered on that adenosine and treated as an independent binary classification instance. A-centered m$^6$A-FORM-sites w/o pretraining used the same adenosine-centered input and sequence-level binary classification formulation but initialized the encoder from scratch.

## Dataset preparation for m$^6$A regulator binding prediction

We collected publicly available CLIP–seq datasets for known m$^6$A regulators (Readers, Writers, Erasers; RWEs) from POSTAR3, which provides high-confidence RBP binding peaks [34]. This resulted in a comprehensive RWE binding dataset comprising 104,244 sequences across 19 regulators, including 13 readers (YTHDF1/2/3, YTHDC1/2, EIF3A/B/D/G, HNRNPA2B1, ELAVL1, IGF2BP1/3), 5 writers (RBM15, RBM15B, WTAP, METTL3, METTL14), and 1 eraser (FTO).

To construct binding labels, we intersected these CLIP-seq peaks with our 131,320 high-confidence base-resolution m$^6$A sites. For each regulator, an m$^6$A site overlapping its CLIP-seq peak was labeled as positive (label = 1), whereas m$^6$A sites without overlap were labeled as negative (label = 0).

For sequence context, labeled sites were mapped to control MeRIP-seq peaks for each RWE. We generated train/validation/test splits at the peak level (70%/10%/20%) using five different random seeds, and extracted 101-nt sequences centered on each labeled site (±50 nt) for each split. To mitigate class imbalance, training and validation used a 1:10 positive-to-negative ratio via random down-sampling of negatives, while the test set preserved the natural class distribution.

## Dataset preparation for YTHDF2-mediated mRNA degradation

To collect m$^6$A site sequences associated with YTHDF2-mediated mRNA degradation, we used YTHDF2 PAR-CLIP data in HeLa cells from GSE49339 [1], consisting of three biological replicates. We retained reproducible YTHDF2-bound genes detected in all three replicates and merged their binding intervals. Coordinates were converted to hg38 using liftOver [56], yielding 52,792 YTHDF2 binding regions. We then intersected these regions with our 131,320 high-confidence base-resolution m$^6$A sites and restricted to HeLa to obtain YTHDF2-bound candidate m$^6$A sites.

To define decay-associated (positive) sites, we incorporated gene-level mRNA half-life measurements in YTHDF2 knockdown (*siYTHDF2*) and control (*siControl*) HeLa cells from [1]. Genes with an average log fold change ($log_2(siYTHDF2/siControl)$) in half-life > 1 were treated as candidate YTHDF2-regulated decay targets. We excluded 5′UTR sites because YTHDF2 binding in 5′UTRs has been reported to preferentially promote translation rather than mRNA degradation [57]. m$^6$A sites overlapping YTHDF2 PAR-CLIP peaks and located within candidate decay genes were labeled as positives, whereas the remaining YTHDF2-bound m$^6$A sites were labeled as negatives.

For sequence context, we mapped positive and negative sites to MeRIP-seq peaks in HeLa control samples, yielding 207 MeRIP-seq peaks. We generated train/validation/test splits at the peak level (70%/10%/20%) using five random seeds, and extracted 501-nt sequences centered on each labeled site (±250 nt) for model training and evaluation.

## Dataset for human tissue MeRIP-seq processing and preparation

Human tissue MeRIP-seq datasets were retrieved from the Chinese Brain Bank Center (CBBC; GSE122744) [47] and the National Genomics Data Center (NGDC; CRA001315) [48]. These datasets comprised 67 samples spanning 24 normal human tissue types, including 22 samples from five male donors in CBBC and 45 samples from seven donors in NGDC. CBBC and NGDC datasets were processed using the same m$^6$A-suite workflow and peak-filtering criteria described above. Briefly, raw reads were quality-controlled, adapter-trimmed, aligned to the GRCh38/hg38 reference genome, and subjected to exomePeak2-based peak calling. High-quality m$^6$A peaks passing the same filtering criteria were used as methylation-enriched input regions for m$^6$A-FORM-sites inference.

## m$^6$A-FORM-RWEBind fine-tuning and evaluation

m$^6$A-FORM-RWEBind was fine-tuned from the pretrained m$^6$A-FORM encoder to predict regulator binding at m$^6$A sites. For each m$^6$A regulator, labeled m$^6$A sites were represented by fixed-length 101-nt sequences centered on the m$^6$A site, corresponding to ±50 nt flanking sequence. Each sequence was tokenized into overlapping 3-mers with a stride of one nucleotide and provided as input to the model. Unlike m$^6$A-FORM-sites, which uses a token-level sequence-labeling formulation, m$^6$A-FORM-RWEBind was formulated as a sequence-level binary classification task. Each input sequence was assigned a positive label if the centered m$^6$A site overlapped the corresponding regulator CLIP-seq peak and a negative label otherwise. This m$^6$A site-centered formulation was used because regulator-binding labels are substantially sparser and less complete than m$^6$A site labels. In sparse-label settings, standard token-level sequence labeling can incorrectly treat unlabeled tokens as true negatives, introducing biased supervision [58].

For each regulator, training, validation, and testing sets were generated using peak-level splits at a ratio of 70:10:20, ensuring that sequences derived from the same MeRIP-seq peak were assigned to the same subset. To reduce the effect of class imbalance during optimization, negative examples in the training and validation sets were randomly downsampled to obtain an approximately 1:10 positive-to-negative ratio. The test set retained the natural positive-to-negative distribution to reflect realistic regulator-binding prediction conditions.

The pretrained m$^6$A-FORM encoder was initialized with the weights learned from MeRIP-seq peak-derived sequences and fine-tuned with a binary classification head. The model generated a continuous binding probability for each m$^6$A site, representing the predicted likelihood that the site is bound by the corresponding regulator. Model selection was performed using validation-set performance, with the best checkpoint selected according to validation PR-AUC. Fine-tuning and evaluation were repeated using five random seeds to account for variation in data splitting and negative sampling.

Model performance was evaluated separately for each regulator using site-level prediction scores. Because regulator-binding labels were highly imbalanced, PR-AUC was used as the primary evaluation metric, and ROC-AUC was also reported. Prediction scores were evaluated as continuous values without applying a fixed threshold for PR-AUC or ROC-AUC calculation. For benchmark comparison, the same training, validation, and test splits were used to train and evaluate baseline RBP-binding prediction models, including iDeepS [23] and RNAProt [24], with input sequences prepared according to each model's required format.

## m$^6$A-FORM-decay fine-tuning and evaluation

m$^6$A-FORM-decay was fine-tuned from the pretrained m$^6$A-FORM encoder to predict m$^6$A sites associated with YTHDF2-mediated mRNA degradation. Labeled YTHDF2-bound m$^6$A sites were represented by fixed-length 501-nt sequences centered on the m$^6$A site, corresponding to ±250 nt flanking sequence. Each sequence was tokenized into overlapping 3-mers with a stride of one nucleotide and provided as input to the model. Similar to m$^6$A-FORM-RWEBind, m$^6$A-FORM-decay was formulated as a sequence-level binary classification task. Among YTHDF2-bound m$^6$A sites, sites located in genes showing increased mRNA half-life upon YTHDF2 knockdown were assigned positive labels, whereas YTHDF2-bound m$^6$A sites without evidence of increased half-life were assigned negative labels.

Training, validation, and test sets were generated using peak-level splits at a ratio of 70:10:20, ensuring that sequences derived from the same MeRIP-seq peak were assigned to the same subset. To reduce the effect of class imbalance during optimization, negative examples in the training and validation sets were randomly downsampled to obtain an approximately 1:1 positive-to-negative ratio. The test set retained the natural positive-to-negative distribution to reflect realistic prediction conditions.

The pretrained m$^6$A-FORM encoder was initialized with the weights learned from MeRIP-seq peak-derived sequences and fine-tuned with a binary classification head.

Model performance was evaluated at the site level using continuous prediction scores. Because decay-associated sites represented a minority of YTHDF2-bound m$^6$A sites, PR-AUC was used as the primary evaluation metric, and ROC-AUC was also reported. Prediction scores were evaluated as continuous values without applying a fixed threshold for PR-AUC or ROC-AUC calculation. For benchmark comparison, m$^6$ABERT-decay [54] was trained and evaluated using the same training, validation, and test splits, with input sequences prepared according to the same ±250 nt site-centered window setting.

## Integrated Gradients and high-attribution region identification

Integrated Gradients (IG) was used to interpret sequence features contributing to m$^6$A-FORM-sites predictions. For each input sequence, IG attribution scores were calculated with respect to the positive m$^6$A prediction score. Given an input token sequence (x) and a reference sequence $x_{\text{ref}}$, IG was computed along a straight-line path from $x_{\text{ref}}$

to $x$:

$$IG(x) = (x - x_{\text{ref}}) \times \int_0^1 \frac{\partial F\left(x_{\text{ref}} + \alpha(x - x_{\text{ref}})\right)}{\partial x} d,$$

where $F(\cdot)$ denotes the model output score and $\alpha$ is the interpolation parameter. The reference input was constructed as a sequence of [PAD] tokens, with [CLS] and [SEP] tokens retained at the beginning and end of the sequence, respectively. Attribution scores were computed on the embedding layer using Captum [59].

Because m$^6$A-FORM-sites uses overlapping k-mer tokens as input, token-level attribution scores were projected back to nucleotide resolution. For each nucleotide, attribution scores from all overlapping k-mers covering that nucleotide were averaged according to the number of contributing tokens. This produced a nucleotide-level attribution profile for each input sequence. Attribution profiles were centered on the predicted m$^6$A site and summarized separately for TC and infrequent m$^6$A sites to compare sequence features prioritized by the model.

High-attribution regions were identified using a strategy similar to that used in DNABERT [53]. Briefly, nucleotide-level attribution profiles were scanned to identify contiguous regions with elevated attribution scores within each input sequence. A region was retained as a high-attribution region if it satisfied the following criteria: (i) each position had an attribution score greater than the mean attribution score of the corresponding sequence; (ii) the attribution score was greater than ten times the minimum attribution score of the corresponding sequence; and (iii) the contiguous region had a minimum length of 4 nt. The resulting high-attribution regions were used for downstream analyses of positional preference, motif enrichment, and overlap with RBP-binding annotations.

## RBP co-binding, occupancy and enrichment analysis

RBP-binding peaks were obtained from POSTAR3 and mapped to hg38 genomic coordinates when necessary. For regulator-specific interpretation analyses, high-attribution regions identified from Integrated Gradients profiles were intersected with RBP-binding peaks using genomic-coordinate overlap analysis. For each candidate co-binding RBP, the number of high-attribution regions overlapping its binding sites was counted separately in the positive and negative groups. Enrichment or depletion of RBP binding in positive high-attribution regions relative to negative high-attribution regions was assessed using Fisher's exact test. P values were adjusted for multiple testing using the Benjamini–Hochberg procedure. To focus on robust enrichment signals, RBPs were retained if they satisfied FDR $< 0.05$, absolute log2 odds ratio $> \log2(1.5)$, and absolute overlap-fraction difference $> 0.01$ between the positive and negative groups.

For tissue-level analyses, predicted tissue m$^6$A sites were intersected with POSTAR3 RBP-binding annotations to quantify local RBP occupancy. The number of unique RBPs overlapping each m$^6$A site was calculated and compared between tissue-conserved and infrequent m$^6$A sites using a two-sided Wilcoxon rank-sum test. For YTHDF2-associated and decay-associated subsets, RBP overlap profiles were compared between tissue-conserved and infrequent sites using Fisher's exact test followed by Benjamini–Hochberg correction. For tissue-level RBP enrichment, RBPs were retained using FDR < 0.05, absolute log2 odds ratio > log2(1.5), and absolute overlap-fraction difference > 0.05.

RBPs enriched near tissue-conserved YTHDF2-associated or decay-associated $m^6A$ sites were used for downstream Gene Ontology Biological Process enrichment analysis using oncoEnrichR [60].

# Identification of tissue-conserved and infrequent $m^6A$ sites

We applied $m^6A$-FORM-sites to $m^6A$ peak regions from 67 samples (22 samples from CBBC and 45 samples from NGDC) identified by exomePeak2. For each dataset, single-base $m^6A$ predictions were filtered using a predefined confidence threshold of 0.78, and sites with prediction scores above this threshold were retained for downstream analysis. High-quality peaks were retained using the peak-filtering criteria described in the pretraining data preparation section.

To account for technical variation in sequencing depth and sample quality, the CBBC and NGDC datasets were analyzed independently. For each dataset, retained $m^6A$ sites from all samples were merged into a non-redundant union set based on genomic coordinates and strand information. A binary site-by-sample incidence matrix was then constructed, in which each row represented a unique predicted $m^6A$ site and each column represented a sample. A value of 1 indicated that a site overlapped a predicted $m^6A$ site in the corresponding sample, whereas a value of 0 indicated absence. The recurrence frequency of each site was calculated as the number of samples in which the site was detected.

To assess whether the observed recurrence frequency exceeded random expectation, we performed a permutation-based recurrence test separately for the CBBC and NGDC datasets. In each permutation, the binary detection labels were randomly shuffled within each sample, thereby preserving the total number of predicted $m^6A$ sites per sample while disrupting site-specific recurrence patterns. This procedure was repeated 1,000 times to generate a null distribution of recurrence frequencies. For each site, an empirical one-sided P value was calculated as the probability that a recurrence frequency from the null distribution was greater than or equal to the observed recurrence frequency. Empirical P values equal to zero were replaced with a pseudocount of 1 divided by the total number of null recurrence values. P values were adjusted for multiple testing using the Benjamini–Hochberg procedure, and sites with adjusted $P < 0.05$ were considered significantly recurrent.

Tissue-conserved $m^6A$ sites were defined as sites that showed significant non-random recurrence in both the CBBC and NGDC datasets. Infrequent $m^6A$ sites were defined as predicted $m^6A$ sites detected in three or fewer tissue types after merging replicate samples from the same tissue within each dataset; sites with intermediate recurrence were excluded from TC-versus-infrequent comparisons.

# Metagene, transcript-region annotation, and GLORI cluster analysis

To examine the transcriptomic distribution of predicted $m^6A$ sites, genomic annotations were obtained from the hg38 transcript database. Coding sequences, 5′ untranslated regions (5′ UTRs), and 3′ untranslated regions (3′ UTRs) were extracted using R package GenomicFeatures [61]. Predicted m$^6$A sites were intersected with each transcript region using genomic-coordinate overlap analysis. The fraction of sites assigned to each region was calculated as the number of unique predicted m$^6$A sites overlapping that region divided by the total number of predicted m$^6$A sites.

Metagene profiles were generated to compare the transcript-level distribution of tissue-conserved and infrequent m$^6$A sites. Tissue-conserved and infrequent m$^6$A sites were represented as GRanges objects and provided as separate groups to GuitarLite. Metagene plots were generated for tissue-conserved and infrequent sites to assess whether the two classes showed distinct positional preferences along mRNA transcripts.

To evaluate whether tissue-conserved m$^6$A sites were associated with clustered m$^6$A organization, we compared them with m$^6$A sites annotated with cluster information from GLORI [15]. GLORI m$^6$A sites from HEK293T cells were imported and converted into single-base GRanges objects using genomic coordinates and strand information. Sites annotated as “Clustered” or “Non-cluster” were separated according to the GLORI cluster annotation. Tissue-conserved and infrequent m$^6$A sites were then intersected with clustered and non-clustered GLORI sites using strand-aware genomic overlap analysis. Enrichment of clustered GLORI sites among tissue-conserved m$^6$A sites relative to infrequent sites was assessed using Fisher’s exact test.

## Methylation level and gene expression association analysis

To compare methylation levels between TC and infrequent m$^6$A sites, sample-specific m$^6$A methylation annotations were obtained for the CBBC and NGDC datasets. Each sample-specific annotation contained genomic coordinates of detected m$^6$A sites together with the estimated methylation level and the corresponding gene expression value derived from the INPUT of MeRIP-seq. TC and infrequent m$^6$A sites were intersected with the sample-specific methylation annotations using genomic-coordinate overlap analysis. For each overlapping site in each sample, the methylation level and gene expression value were extracted and assigned to the corresponding predicted m$^6$A site.

For each site, methylation levels across all available CBBC and NGDC samples were aggregated by calculating the mean methylation level across samples with non-missing values. The aggregated methylation level was then transformed as log2(methylation level + 1).

To evaluate the relationship between m$^6$A methylation level and host gene expression, gene expression values associated with each m$^6$A site were extracted from the same sample-specific methylation annotations. For each TC m$^6$A site, gene expression values were averaged across samples with available measurements and transformed as log2(expression + 1). The association between the aggregated methylation level and gene expression was then assessed using Spearman’s rank correlation.

## Prediction of single-base m$^6$A sites, YTHDF2 binding potential, and decay propensity

To characterize the regulatory potential of tissue m$^6$A sites, we applied the m$^6$A-FORM framework to predict single-base m$^6$A sites, YTHDF2 binding potential, and mRNA decay propensity. For each tissue sample, high-quality m$^6$A peak regions identified from MeRIP-seq data were used as input regions. Candidate adenosines within these peak regions were extracted according to the hg38 reference genome, and peak-derived sequences were tokenized into overlapping 3-mers. $m^6A$-FORM-sites was then applied to predict high-confidence single-base $m^6A$ sites within these regions.

Predicted $m^6A$ sites were subsequently used for YTHDF2-binding and decay-propensity prediction. For YTHDF2 binding prediction, fixed-length sequence windows of ±50 bp centered on each predicted $m^6A$ site were generated, tokenized into overlapping 3-mers, and provided as input to $m^6A$-FORM-RWEBind. Sites with YTHDF2-binding prediction scores > 0.76 were retained as candidate YTHDF2-binding $m^6A$ sites.

To estimate the potential effect of YTHDF2-associated $m^6A$ sites on mRNA degradation, candidate YTHDF2-binding $m^6A$ sites were further analyzed using $m^6A$-FORM-decay. For each site, a fixed-length sequence window of ±250 bp centered on the predicted $m^6A$ site was generated and used as input to $m^6A$-FORM-decay to obtain a decay-propensity score. This score represented the predicted likelihood that the $m^6A$ site is associated with YTHDF2-mediated mRNA degradation. Sites with decay-propensity scores > 0.42 were retained as candidate decay-associated $m^6A$ sites. Candidate sites overlapping annotated 5′ UTRs were removed before downstream analysis.

## Funding Statement

Y. H. discloses support for the research of this work from National Institutes of Health [U01CA279618 and R21GM155774]. S.-J. G. discloses support for the research of this work from National Institutes of Health [CA096512, CA284554, CA278812, CA291244 and CA124332]. Y. H. and S. -J. G disclose support for research of this work from UPMC Hillman Cancer Center Startup Fund and in part by award [P30CA047904]. This research was supported in part by the University of Pittsburgh Center for Research Computing through the resources provided. Specifically, this work used the HTC cluster, which is supported by National Institutes of Health award [S10OD028483].

## Data and code availability

The corresponding data are available on GitHub at https://github.com/Huang-AI4Medicine-Lab/TC-$m^6A$; Code is available on GitHub at https://github.com/TingheZhang/$m^6A$-FORM.

## Materials & Correspondence.

Correspondence and material requests should be addressed to Dr. Yufei Huang (yuh119@pitt.edu)

# Rerferences

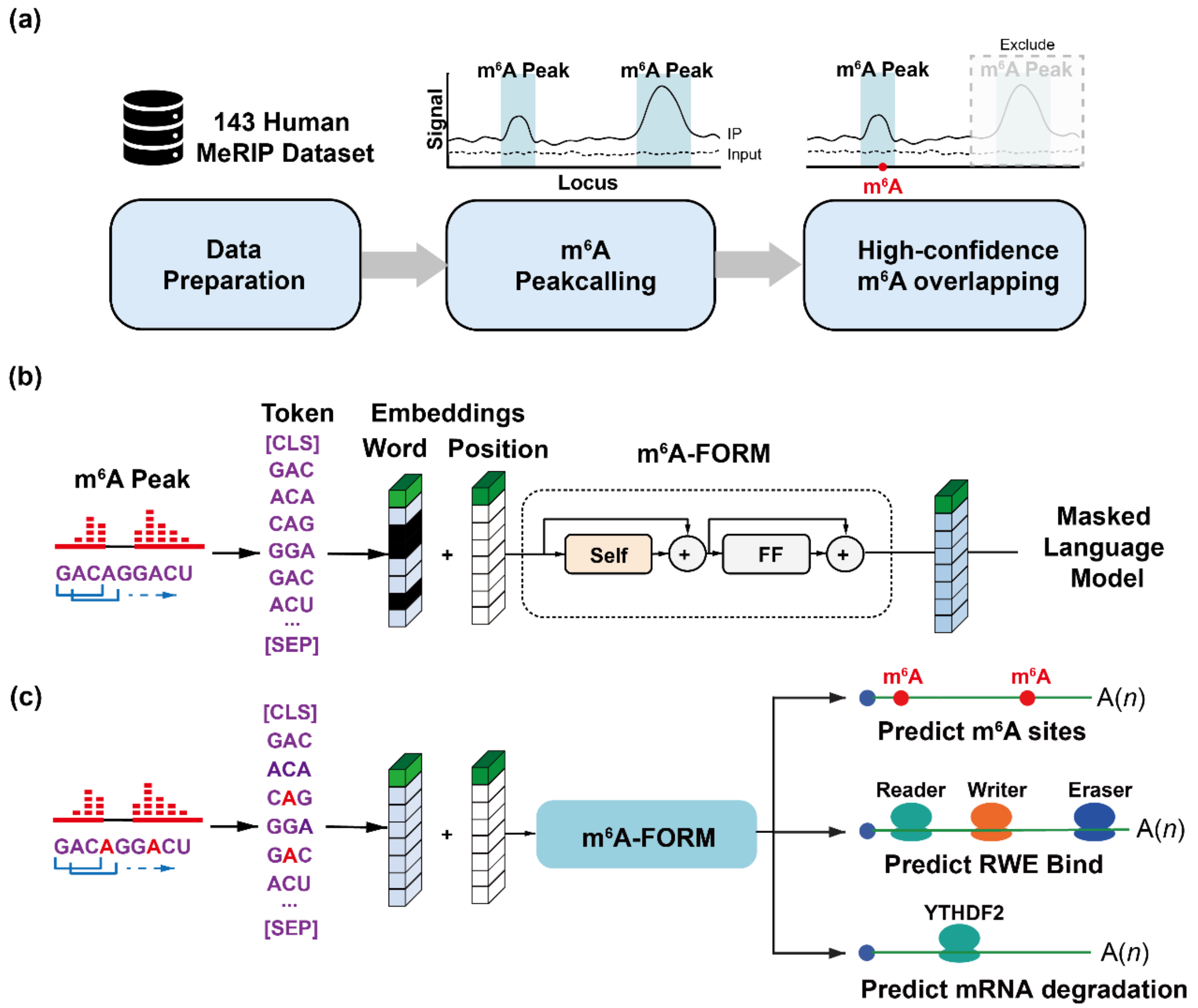

Fig. 1 | Overview of the m$^6$A-FORM framework.

a, Pipeline for constructing the high-confidence single-base m$^6$A dataset. A total of 143 human MeRIP-seq datasets were processed through data preparation and peak calling, yielding 24,909,934 mRNA sequences derived from 22,548,379 MeRIP-seq peaks for pretraining. High-confidence single-base m$^6$A sites overlapping called peaks were subsequently used as ground truth for downstream tasks.

b, Architecture of the m$^6$A-FORM model. RNA sequences within m$^6$A peaks were tokenized as 3-mers and embedded before being passed through 12 Transformer blocks. The model was pretrained using a masked language modelling objective to capture contextual features associated with m$^6$A.

c, Downstream tasks enabled by m$^6$A-FORM. Following pretraining, the model was applied to m$^6$A-site prediction, reader/writer/eraser (RWE) protein-binding prediction and YTHDF2-mediated mRNA degradation prediction, demonstrating its capacity to learn biologically meaningful regulatory signals across diverse m$^6$A-related tasks.

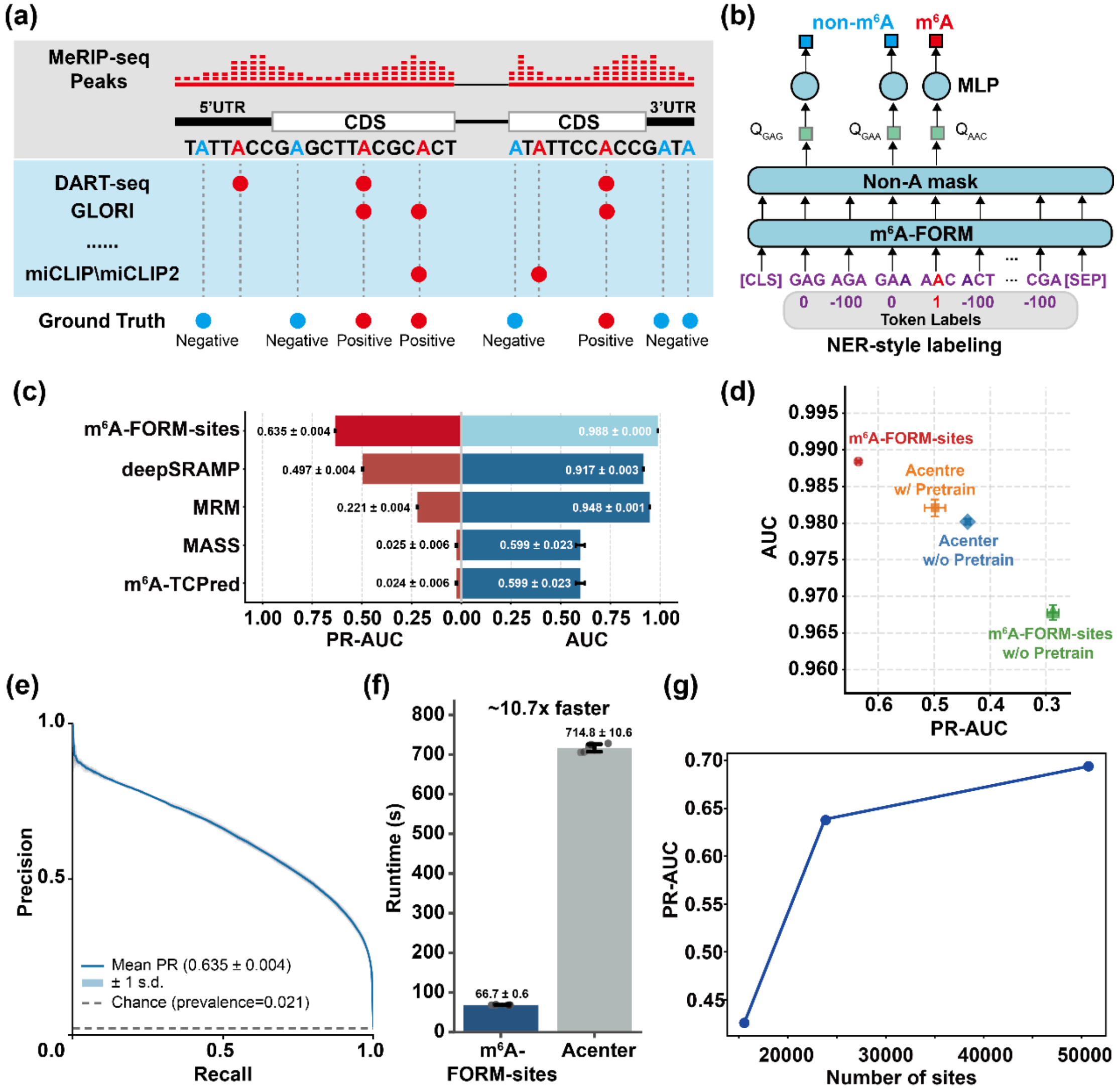

Fig. 2 | Construction of high-confidence m$^6$A ground truth and evaluation of m$^6$A-FORM-sites.

a, Schematic of high-confidence single-base m$^6$A ground-truth generation. Data from 13 single-base-resolution m$^6$A sequencing technologies were collected from m$^6$A-Atlas V2 and GLORI. Adenosines supported by at least two independent technologies were labelled as positives, whereas unsupported adenosines were labelled as negatives. Sites identified by only one technology were excluded from training to minimize label noise. Because miCLIP and miCLIP2 rely on similar experimental principles, they were treated as a single technology.

b, Non-A masking layer and NER-style labelling strategy for m$^6$A-site prediction. Non-adenosine residues were masked, and the filtered representations were fed into a shared multilayer perceptron (MLP). For each 3-mer token, the central nucleotide was used for labelling: 0 denotes a non-methylated adenosine, 1 denotes an m$^6$A site and −100 denotes ignored positions. This strategy enables simultaneous prediction of all candidate adenosines within a single input sequence.

c, Performance benchmarking against state-of-the-art baseline models. m$^6$A-FORM-sites outperformed competing models in both PR-AUC and ROC-AUC. Values are shown as mean ± s.d. across five random splits.

d, Ablation analysis of the effects of pretraining and labelling strategy. The two-dimensional scatter plot with error bars shows that both NER-style labelling and large-scale pretraining contributed to model performance.

e, Precision–recall curves for $m^6A$-FORM-sites. Values are shown as mean ± s.d. across five random splits.

f, GPU runtime comparison. $m^6A$-FORM-sites with NER-style labelling was approximately 10.7-fold faster than the A-centred labelling method.

g, Scaling analysis showing the improvement in PR-AUC as the amount of high-confidence ground-truth training data increased. Relaxing the labelling criteria increased the number of positive sites and led to gains in PR-AUC, suggesting that $m^6A$-FORM-sites benefits from expanded training data while remaining sensitive to the precision–coverage trade-off associated with label confidence.

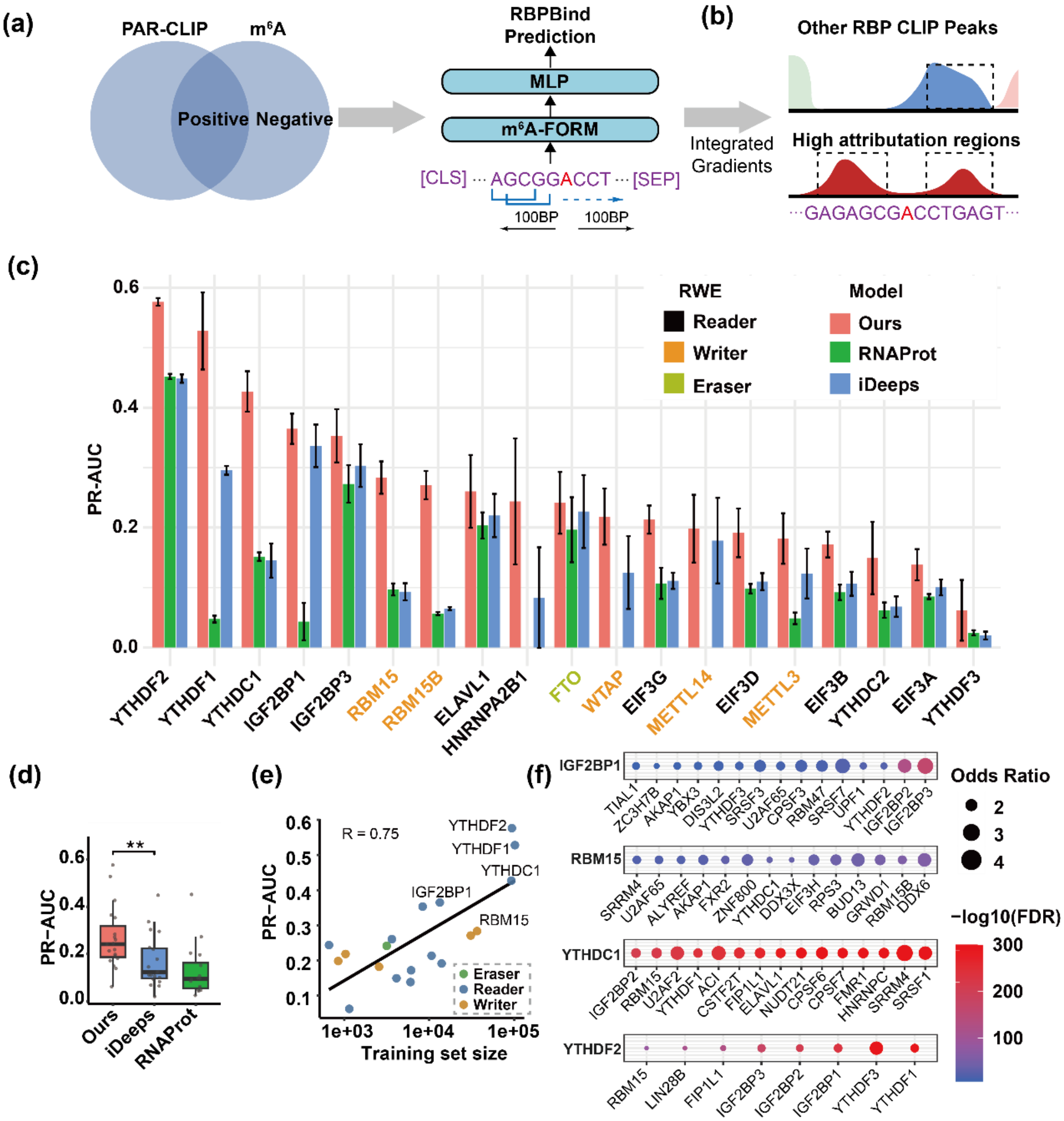

Fig. 3 | m$^6$A-FORM-RWEBind predicts m$^6$A-regulator binding and identifies candidate RBP cofactors.

a, Schematic of the m$^6$A-FORM-RWEBind framework. CLIP-seq peaks of m$^6$A readers, writers and erasers were intersected with high-confidence single-nucleotide m$^6$A sites to define regulator-specific positive and negative binding events. Fixed A-centred sequence windows were encoded by m$^6$A-FORM and fine-tuned with an MLP head for regulator-binding prediction.

b, Attribution-guided cofactor discovery for selected m$^6$A regulators. Integrated gradients were used to identify high-attribution sequence regions for each trained regulator-specific model, which were then intersected with POSTAR3 RBP CLIP-seq peaks. RBPs enriched in high-attribution regions from positive sites relative to negative sites were selected as candidate cofactors.

c, PR-AUC comparison of m$^6$A-FORM-RWEBind with benchmark models across 19 m$^6$A regulators. Regulator labels are coloured by functional class: readers, writers and erasers. Error bars indicate variation across five random repeats.

d, Distribution of mean PR-AUC across regulators for each model. Box plots summarize regulator-level performance distributions. Statistical significance was assessed using a paired test comparing $m^6A$-FORM-RWEBind with the second-best model.

e, Relationship between regulator-specific training-set size and PR-AUC. Each point represents one $m^6A$ regulator.

f, Top candidate RBP cofactors identified by attribution-guided enrichment analysis for selected $m^6A$ regulators. Enriched RBPs were ranked by significance.

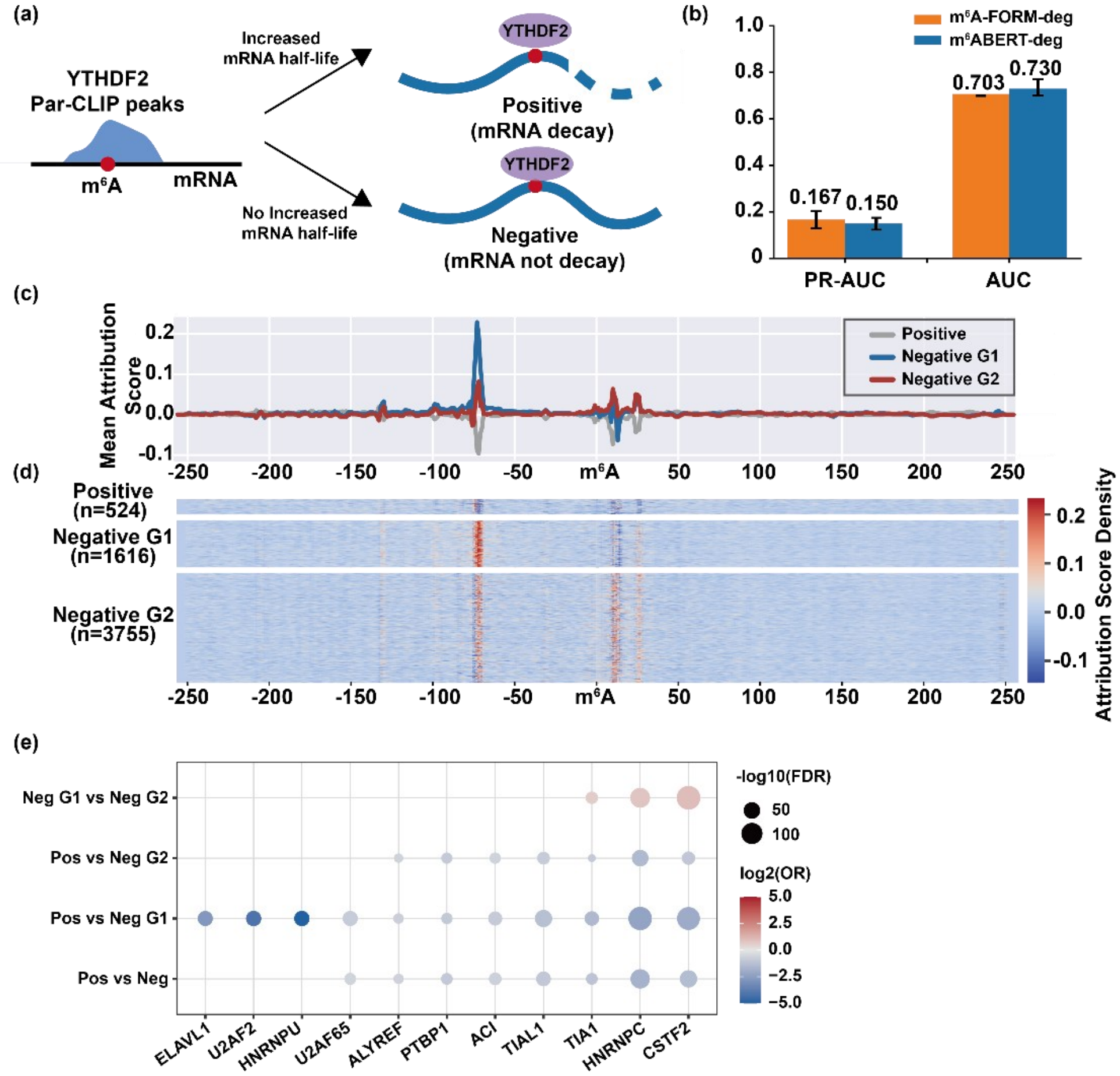

Fig. 4 | Prediction and interpretation of YTHDF2-associated mRNA degradation using m$^6$A-FORM-decay.

a, Schematic of the ground-truth labelling strategy for degradation-associated m$^6$A sites. Candidate m$^6$A sites overlapping YTHDF2 PAR-CLIP peaks were labelled as positives when located in transcripts showing increased mRNA half-life upon YTHDF2 knockdown, whereas sites in transcripts without increased half-life were labelled as negatives.

b, Model performance in identifying degradation-associated m$^6$A sites. m$^6$A-FORM-decay achieved comparable or improved performance relative to m$^6$ABERT-deg, with PR-AUC and AUC values of 0.167 and 0.703, respectively.

c,d, Integrated gradients-based interpretation analysis around m$^6$A sites. Mean attribution profiles (c) and site-level attribution heatmaps (d) revealed distinct attribution patterns between positive and negative sites. Negative sites were further separated into two subgroups, indicating heterogeneous sequence-context features among non-degradation-associated m$^6$A sites.

e, RBP enrichment analysis comparing positive sites with the two negative subgroups. Subgroup-specific enrichment patterns indicate that different classes of YTHDF2-bound m$^6$A sites are associated with distinct local RBP environments.

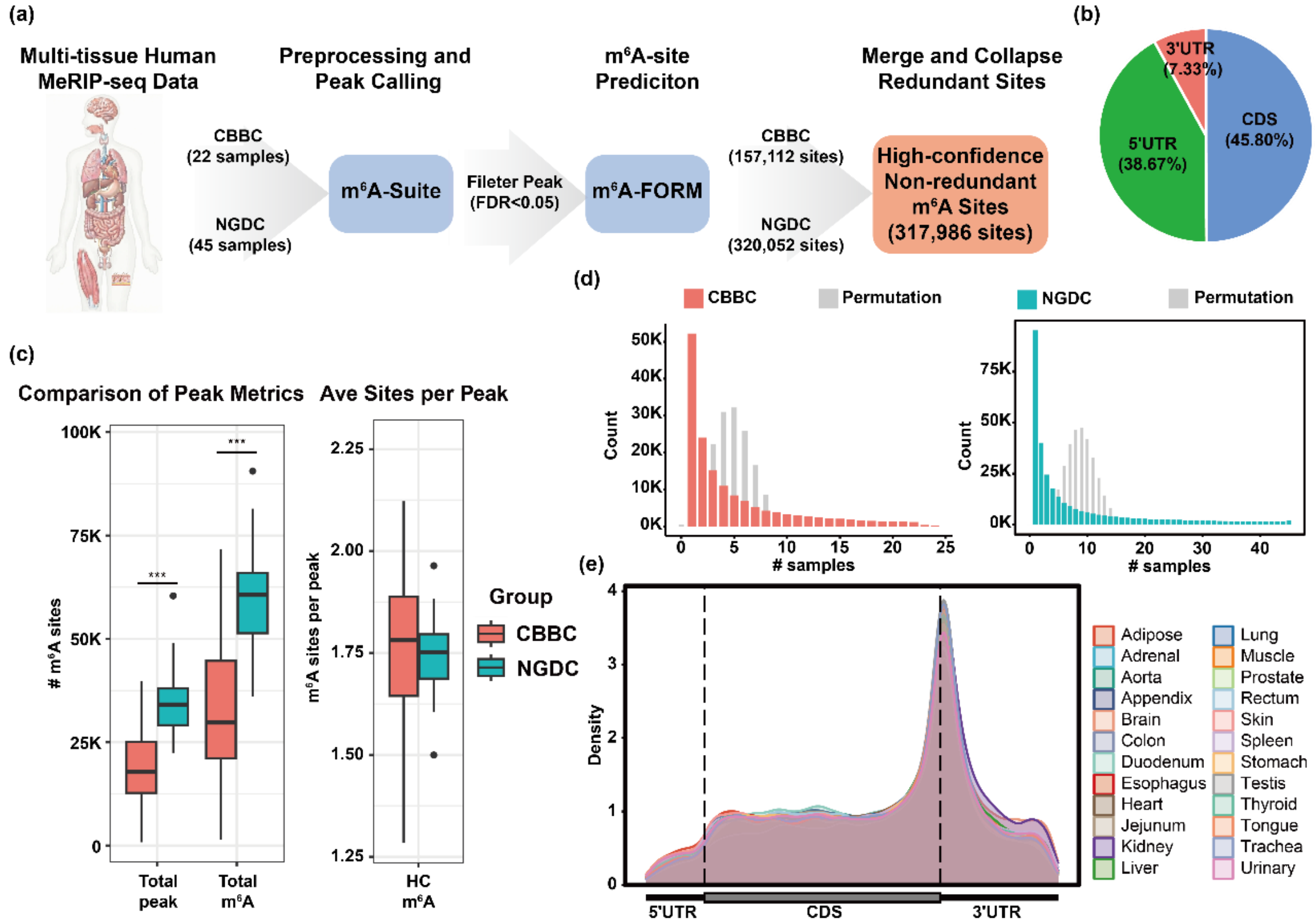

Fig. 5 | Multi-tissue landscape and peak characteristics of high-confidence human m$^6$A sites.

a, Schematic overview of the workflow for deriving high-confidence m$^6$A sites from human multi-tissue MeRIP-seq datasets. MeRIP-seq samples from the CBBC and NGDC were processed for peak calling, followed by site prediction with m$^6$A-FORM and collapsing of redundant sites, yielding 317,986 high-confidence non-redundant m$^6$A sites.

b, Transcript-region distribution of the predicted high-confidence m$^6$A sites.

c, Comparison of peak-level characteristics between the CBBC and NGDC datasets. Although the total numbers of peaks and predicted m$^6$A sites differed between datasets, the average number of m$^6$A sites per peak remained relatively consistent, supporting the robustness of site-level prediction.

d, Distribution of sample recurrence for individual m$^6$A sites in the CBBC and NGDC datasets compared with the corresponding permutation-based null distributions.

e, Metagene distribution of high-confidence m$^6$A sites across 24 human tissues. Predicted sites exhibited a conserved positional pattern along transcripts.

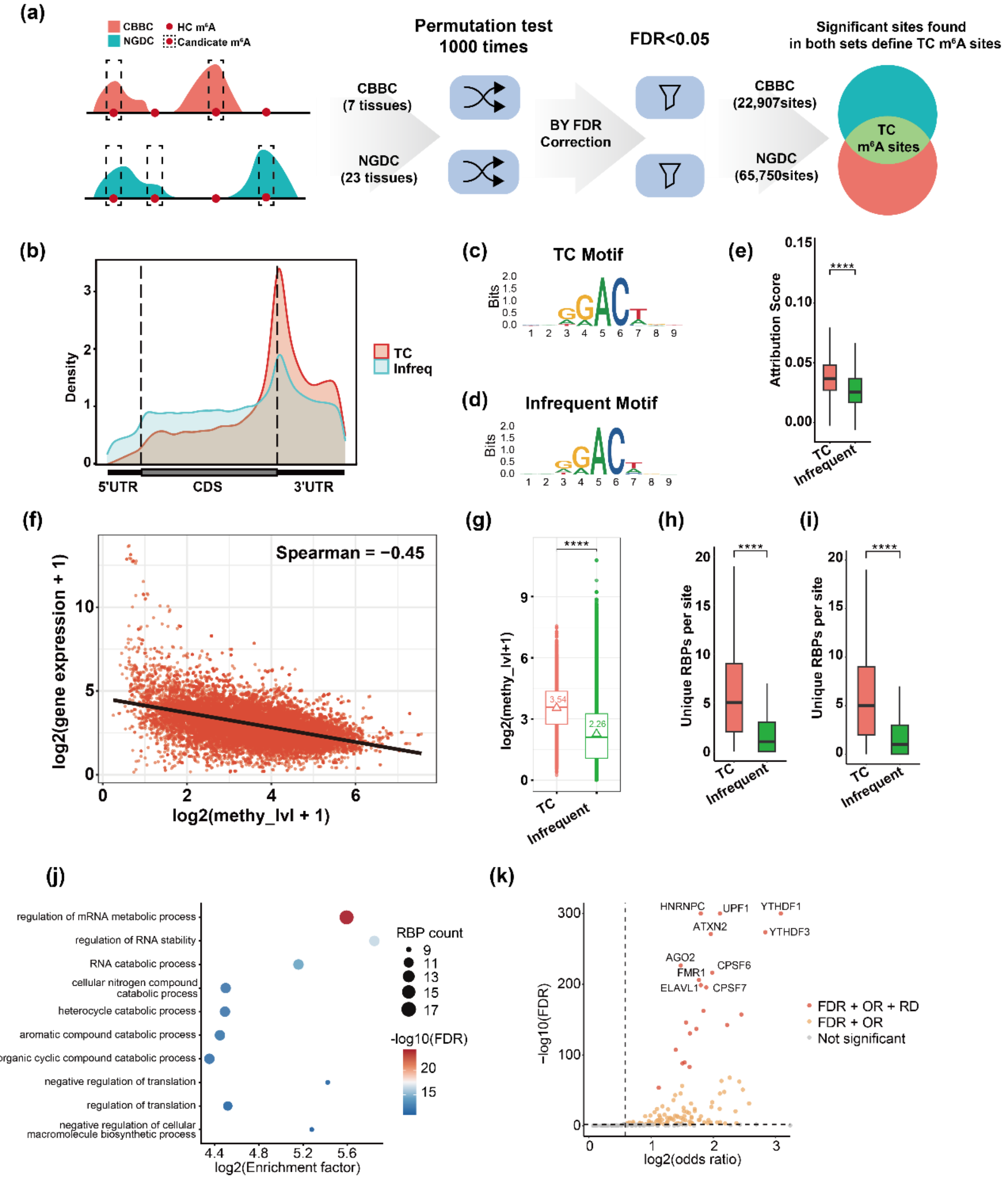

Fig. 6 | Tissue-conserved m$^6$A sites exhibit distinct molecular and regulatory features.

a, Schematic definition of tissue-conserved (TC) and infrequent m$^6$A sites. Candidate TC sites were identified as recurrent m$^6$A sites shared across tissues and supported by both CBBC and NGDC datasets after permutation testing and FDR correction.

b, Metagene distribution of TC and infrequent m$^6$A sites across transcript regions.

c,d, Sequence motifs associated with TC (c) and infrequent (d) m$^6$A sites.

e, Comparison of average attribution scores between TC and infrequent m$^6$A sites across 24 tissues. TC sites showed significantly higher attribution scores than infrequent sites.

f, Relationship between methylation level and gene expression for TC m$^6$A sites.

g, Comparison of methylation levels between TC and infrequent $m^6A$ sites.

h,i, Comparison of local RBP occupancy around TC and infrequent $m^6A$ sites near YTHDF2-binding sites (h) and YTHDF2-mediated mRNA decay sites (i).

j, GO biological process enrichment analysis of RBPs located around $m^6A$ sites with evidence of YTHDF2 binding or YTHDF2-mediated mRNA decay, comparing TC and infrequent sites.

k, Volcano plot showing RBP enrichment around TC versus infrequent $m^6A$ sites. Significant RBPs were identified using Fisher's exact test with combined thresholds for FDR, odds ratio (OR) and ratio difference (RD).